\documentclass[aps,prl,twocolumn,showpacs,citeautoscript,superscriptaddress]{revtex4-1}

\usepackage{graphicx}  % needed for figures
\usepackage{dcolumn}   % needed for some tables
\usepackage{bm}        % for math
\usepackage{amsmath}
\usepackage{amssymb}   % for math
\usepackage{array}
\usepackage[bookmarks=false,linkcolor=blue,urlcolor=blue,colorlinks,citecolor=blue]{hyperref}
\usepackage{exscale,relsize}
\usepackage{color}
\usepackage{times, verbatim}
\usepackage{bbold}
\usepackage{bm}
\usepackage{pst-all}
\usepackage{subfigure}

\newcommand{\bra}[1]{\langle #1|}
\newcommand{\ket}[1]{|#1\rangle}

\makeatletter

\begin{document}

\title{Survival of the fractional Josephson effect in time-reversal-invariant topological superconductors}

\author{Christina Knapp}\thanks{These authors contributed equally to this work.}
\affiliation{Department of Physics and Institute for Quantum Information and Matter, California Institute of Technology, Pasadena,
	California 91125 USA}
\affiliation{Walter Burke Institute for Theoretical Physics, California Institute of Technology, Pasadena,
	California 91125 USA}
\author{Aaron Chew}\thanks{These authors contributed equally to this work.}
\affiliation{Department of Physics and Institute for Quantum Information and Matter, California Institute of Technology, Pasadena,
	California 91125 USA}
\author{Jason Alicea}
\affiliation{Department of Physics and Institute for Quantum Information and Matter, California Institute of Technology, Pasadena,
	California 91125 USA}
\affiliation{Walter Burke Institute for Theoretical Physics, California Institute of Technology, Pasadena,
	California 91125 USA}

\begin{abstract}
Time-reversal-invariant topological superconductor (TRITOPS) wires host Majorana Kramers pairs that have been predicted to mediate a fractional Josephson effect with $4\pi$ periodicity in the superconducting phase difference.
We explore the TRITOPS fractional Josephson effect in the presence of time-dependent `local mixing' perturbations that instantaneously preserve time-reversal symmetry.
Specifically, we show that just as such couplings render braiding of Majorana Kramers pairs non-universal, the Josephson current becomes either aperiodic or $2\pi$-periodic (depending on conditions that we quantify) unless the phase difference is swept sufficiently quickly.
We further analyze topological superconductors with $\mathcal{T}^2 = +1$ time-reversal symmetry and reveal a rich interplay between interactions and local mixing that can be experimentally probed in nanowire arrays.
\end{abstract}

\date{\today}

\maketitle

{\bf \emph{Introduction.}}~Topological superconducting wires hosting unpaired end Majorana zero modes (MZMs) \cite{Kitaev01,Beenakker13,Alicea12,Leijnse12,Sato16,Stanescu13,Elliott15,DasSarma15,Aguado17,Lutchyn17} display remarkably rich phenomenology with applications for topological quantum computation \cite{Kitaev03,Nayak08,Karzig16,Plugge17,Vijay16}.
Well-separated MZMs span a set of degenerate ground states that are locally indistinguishable and hence define a fault-tolerant qubit subspace.
Braiding unpaired MZMs implements universal non-Abelian rotations within the ground state subspace---thereby generating fault-tolerant qubit gates.
Under a `fusion' process, a pair of MZMs brought together in space hybridizes and yields a finite-energy fermionic state that can be either empty ($I$ fusion channel) or filled ($\psi$ fusion channel).
Detecting which fusion channel emerges provides a means of qubit readout.

An elegant method of probing topological superconductivity and performing readout utilizes the so-called fractional Josephson effect~\cite{Kitaev01,Fu07,Fu09,Kwon03}.
In a conventional Josephson junction, Cooper-pair tunneling generates a current that is $2\pi$ periodic in the superconducting phase difference across the barrier.
A pair of MZMs fused across a topological Josephson junction mediates single-electron tunneling, resulting in an anomalous $4\pi$-periodic fractional Josephson current whose sign correlates with the associated fusion channel.
This period doubling intimately relates to non-Abelian braiding:
Advancing the phase difference by $2\pi$ has the same effect as fully braiding the MZM pair on one side of the junction (in turn swapping the fusion channel $I \leftrightarrow \psi$ for the hybridized MZMs).
Two such braids are necessary to return the system to its initial state, corresponding to $4\pi$ phase evolution.

When some time-reversal symmetry $\mathcal{T}$ is present, topological superconducting wires can host multiple MZMs at each boundary.
In particular, a time reversal invariant topological superconductor (TRITOPS) for which $\mathcal{T}^2=-1$~\cite{Qi09,Fu10,Santos10,Deng11,Deng12a,Nakosai12,Wong12,Keselman13,Dumitrescu13,Nakosai13,Zhang13,Liu14,Sun14,Haim14,Klinovaja14,Schrade15,Pikulin16,Haim16,Ebisu16,Kim16,Li16,Alase16,Parhizgar17,Reeg17,Haim18,Aligia18,Schrade18,Casas19}
hosts a Kramers pair of end MZMs that cannot hybridize provided $\mathcal{T}$ is preserved.
MZM Kramers pairs in a TRITOPS wire accordingly generate a symmetry-protected ground state degeneracy consisting of locally distinguishable states---and thus furnish a qubit subspace with limited fault-tolerance.
Indeed, even time-dependent local perturbations that instantaneously preserve $\mathcal{T}$ can rotate the Majorana Kramers pair wavefunctions, generating a non-universal non-Abelian Berry phase~\cite{Wolms14}.
As a result of this `local mixing,' braiding MZM Kramers pairs generically produces {\it non-universal} rotations in the ground-state subspace~\cite{Wolms16}.

\begin{figure}
	\includegraphics[width=\columnwidth]{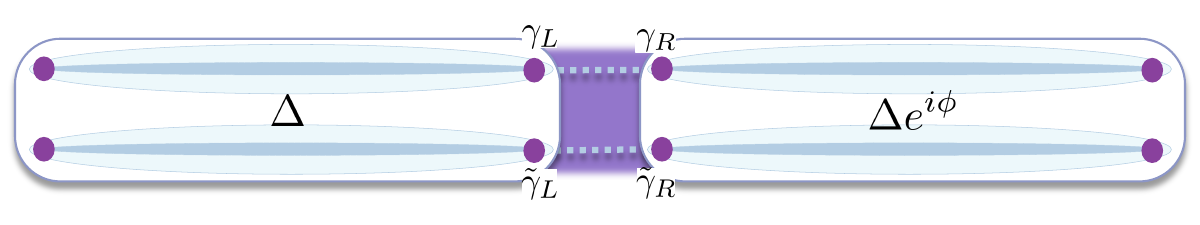}
	\includegraphics[width=.9\columnwidth]{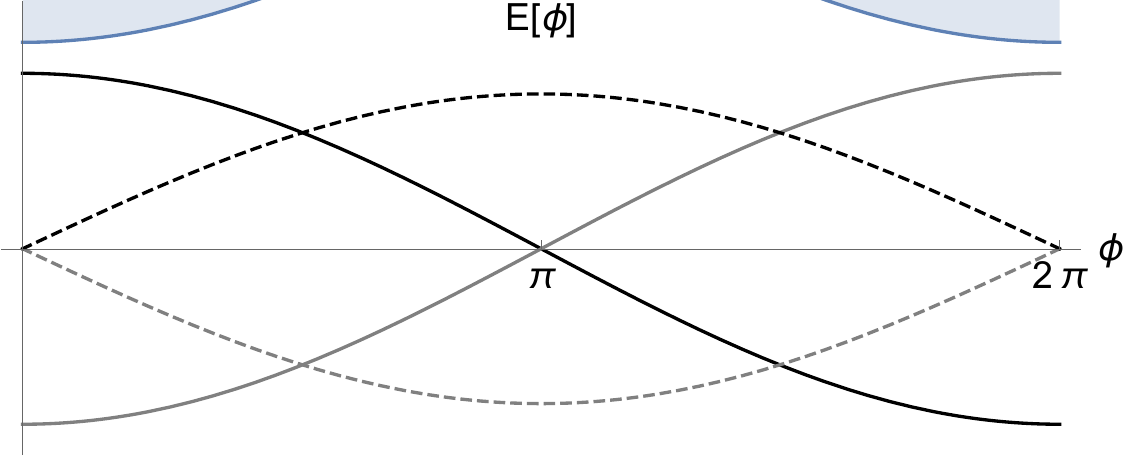}
	\caption{  {\it Top panel:} TRITOPS Josephson junction with each wire modeled by two time-reversed copies of a topological superconductor.
	Dotted lines indicate hybridization of Majorana Kramers pairs (dark purple) leading to Eq.~\eqref{eq:HJJ}.
	Local mixing arises when Majorana Kramers pairs undergo time-dependent coupling to excited states.
	{\it Bottom panel:} Many-body energy spectrum for a TRITOPS Josephson junction.  Solid and dashed curves respectively correspond to even and odd fermion-parity sectors.
}\label{fig:Tm1}
\end{figure}

Given these non-universalities, to what extent does a fractional Josephson effect survive in TRITOPS wires?
This question turns out to be exceedingly subtle.
On one hand, in a TRITOPS Josephson junction that preserves $\mathcal{T}$ at phase differences $0$ and $\pi$, each subgap level is certainly $4\pi$-periodic (Fig.~\ref{fig:Tm1}), suggesting that a fractional Josephson effect appears as predicted in numerous works~\cite{Zhang13,Chung13,Keselman13,Liu14,Zhang14b,Mellars16,Gong16,Liu16,Camjayi17,Alase17,Haim18,Cobanera18,Arrachea19}.
But on the other, the braiding/fractional Josephson connection noted earlier naively implies that non-universality of the former spells doom for the latter.
There is, however, reason for optimism: The Josephson-junction energy levels become degenerate only at {\it discrete} phase differences, suggesting that time-dependent local perturbations may play a less dramatic role compared to the braiding problem (for which degeneracy persists throughout the evolution).

Here we show that, when the superconducting phase winds adiabatically, local mixing indeed spoils the fractional Josephson effect and yields either an aperiodic or $2\pi$-periodic current-phase relation depending on local-mixing time scales.
This result holds even in an otherwise ideal situation for which effects known previously to destroy $4\pi$-periodicity~\cite{Fu09,SanJose12,Pikulin11,Lee14,Chung13,Badiane13,Mellars16,Peng16,Sticlet18}---e.g., explicit $\mathcal{T}$ breaking, overlap between distant MZMs, energy relaxation, and quasiparticle poisoning---are absent.
By mapping the problem onto an effective model that features avoided crossings in the energy spectrum, we further demonstrate that $4\pi$ periodicity is recovered when the phase difference evolves sufficiently quickly that local mixing remains benign.
We extend our analysis to junctions of $\mathcal{T}^2=+1$ topological superconductors, which can be realized (approximately) with proximitized nanowire arrays \cite{Tewari12}.
Without interactions, local mixing can similarly spoil the fractional Josephson effect in a junction of $2m$ wires when $m>1$.
Interestingly, however, we find that interactions stabilize $4\pi$ periodicity for any odd $m$ and $2\pi$ periodicity when $m\mod 4= 0$.
The nontrivial $m$ dependence reflects an interplay between local mixing and the $\mathbb{Z}_8$ classification of one-dimensional fermionic topological phases~\cite{Fidkowski09}, and thus provides an experimental window into both phenomena.

%%%
{\bf \emph{Local mixing.}}~Let $\gamma_{1(2)},$ $\tilde{\gamma}_{1(2)}$ denote the MZM Kramers pair at the left (right) end of a TRITOPS wire.
Time reversal sends
\begin{align}
  \gamma_j \rightarrow s_j \tilde \gamma_j,~~~ \tilde \gamma_j \rightarrow - s_j\gamma_j
  \label{eq:TRS-gamma}
\end{align}
for convention-dependent signs $s_j = \pm 1$; note consistency with $\mathcal{T}^2 = -1$.
Bilinears hybridizing a given MZM Kramers pair are odd under $\mathcal{T}$ and thus forbidden.

Local, adiabatic time-dependent perturbations that instantaneously preserve $\mathcal{T}$ endow the MZM operators at each end with nontrivial time dependence.
Such perturbations can result, e.g., from external manipulation or stochastic noise that couples the initial MZMs and bulk energy modes, as reviewed in the Supplemental Material.
Reference~\onlinecite{Wolms14} showed that after the Hamiltonian completes a closed cycle in time $T$, the final state generically differs in a non-universal way from the initial state.
The ground-state rotation resulting from this ``local mixing" is implemented by the unitary matrix ${U = \exp \left[\sum_{j = 1,2}\theta_j \gamma_j(0)\tilde{\gamma}_j(0) /2 \right]}$, with $\theta_j$ a local-mixing angle determined by evolution details.
Local mixing accordingly spoils the topological protection of braiding MZM Kramers pairs~\cite{Wolms16}.

{\bf \emph{TRITOPS Josephson junction.}}~Consider a TRITOPS Josephson junction (Fig.~\ref{fig:Tm1}, top) with superconducting phase $\phi_L = 0$ on the left and $\phi_R = \phi$ on the right.
The Majorana Kramers pair $\gamma_L, \tilde \gamma_L$ at the left side of the junction hybridizes with the Majorana Kramers pair $\gamma_R, \tilde \gamma_R$ on the right side, mediating a supercurrent contribution that we wish to explore in the presence of local mixing.
When $\phi = n \pi$ for $n \in \mathbb{Z}$ the junction preserves $\mathcal{T}$;
at these values we adopt a convention where $\gamma_L, \tilde \gamma_L$ transform under $\mathcal{T}$ according to Eq.~\eqref{eq:TRS-gamma} with $s_L = 1$, while $\gamma_R, \tilde \gamma_R$ transforms with $s_R = -(-1)^n$.
A minimal time-independent junction Hamiltonian compatible with this symmetry reads
\begin{align}\label{eq:HJJ}
H_\text{JJ} &= i \lambda_e\cos\left(\phi/2\right)\left(\gamma_{L}\gamma_{R}+\tilde{\gamma}_L\tilde{\gamma}_R \right)
\nonumber \\
&~~~~~+ i \lambda_o \sin\left(\phi/2\right)\left(\gamma_{L}\gamma_{R}-\tilde{\gamma}_L\tilde{\gamma}_R \right),
\end{align}
where $\lambda_{e,o}$ are real-valued tunneling amplitudes (see Supplemental Material for a derivation).
In the even-parity sector $({i\gamma_L\gamma_R)(i\tilde{\gamma}_L\tilde{\gamma}_R)} = +1$ only the first line survives, whereas in the odd-parity sector $({i\gamma_L\gamma_R)(i\tilde{\gamma}_L\tilde{\gamma}_R)} = -1$ only the second line survives
\footnote{Interactions generate a four-body term $\propto \gamma_L\gamma_R\tilde{\gamma}_L\tilde{\gamma}_R$ that can further split the even- and odd-parity sectors, but do not play an important role in this setup.}.
Figure~\ref{fig:Tm1}, bottom panel, sketches the corresponding energy-phase relation $E(\phi)$.
Each subgap energy is $4\pi$-periodic in $\phi$, and hence at this level of analysis the Josephson current $\langle I(\phi) \rangle = \frac{2e}{\hbar} \frac{d E}{d\phi}$ is also $4\pi$-periodic under adiabatic phase evolution.

Crucially, however, the TRITOPS fractional Josephson effect hinted at here does {\it not} constitute a robust adiabatic cycle.
As proof of concept, suppose that we begin in the ground state of the even-parity sector (solid lines in Fig.~\ref{fig:Tm1}, bottom), and then implement the following process:
$(i)$ adiabatically wind $\phi$ from $0$ to $\pi$, yielding two-fold Kramers degeneracy, $(ii)$ turn on a local-mixing closed adiabatic subcycle, and $(iii)$ adiabatically wind $\phi$ from $\pi$ to $2\pi$.
After stage $(ii)$ local mixing rotates the system into a superposition of even-parity junction eigenstates via ${U_{\rm JJ} = \exp \left[\sum_{J = L,R}\theta_J \gamma_J(0)\tilde{\gamma}_J(0) /2 \right]}$ for some non-universal $\theta_{L,R}$ mixing angles.
Specifically, if $\ket{-}$ and $\ket{+}$ denote states that respectively evolve from the even-parity ground state and excited state at $\phi = \pi$, the system evolves to
\begin{align}\label{eq:UJJ}
U_{\rm JJ}\ket{-}&= \cos\left(\frac{\theta_R-\theta_L}{2}\right)\ket{-} +i \sin\left(\frac{\theta_R-\theta_L}{2}\right) \ket{+}.
\end{align}
Repeated implementations of the closed adiabatic cycle above generically result in {\it aperiodic} unitary state evolution, signaling a breakdown of the TRITOPS fractional Josephson effect.

A more physically relevant scenario arises when local mixing and phase winding occur simultaneously.
For an illustrative toy model, we incorporate a Kramers pair of Andreev bound states described by $f = (\gamma'_\varepsilon + i\gamma_\varepsilon)/2, \tilde f = (\tilde \gamma'_\varepsilon + i\tilde \gamma_\varepsilon)/2$ and supplement Eq.~\eqref{eq:HJJ} with
\cite{Wolms14}
\begin{align}\label{eq:HLM}
&\delta H(t) = \frac{\varepsilon}{2} \left[-i\left( \gamma_\varepsilon \gamma_{\varepsilon}' + \tilde{\gamma}_{\varepsilon}\tilde{\gamma}_{\varepsilon}'\right) +2 \right]
\\
& + i  \frac{\beta}{2} \left[ \cos\alpha(t) \left( \gamma_L \gamma_{\varepsilon} + \tilde{\gamma}_L \tilde{\gamma}_{\varepsilon} \right) + \sin\alpha(t) \left( \gamma_L \tilde{\gamma}_\varepsilon - \tilde{\gamma}_L \gamma_{\varepsilon} \right) \right] .
\nonumber
\end{align}
The Andreev bound states exhibit an energy gap $\varepsilon$ encoded by the first line and couple to the Majorana Kramers pair on the left side of the junction via the second line; all terms instantaneously preserve $\mathcal{T}$.

\begin{figure}
	\includegraphics[width=\columnwidth]{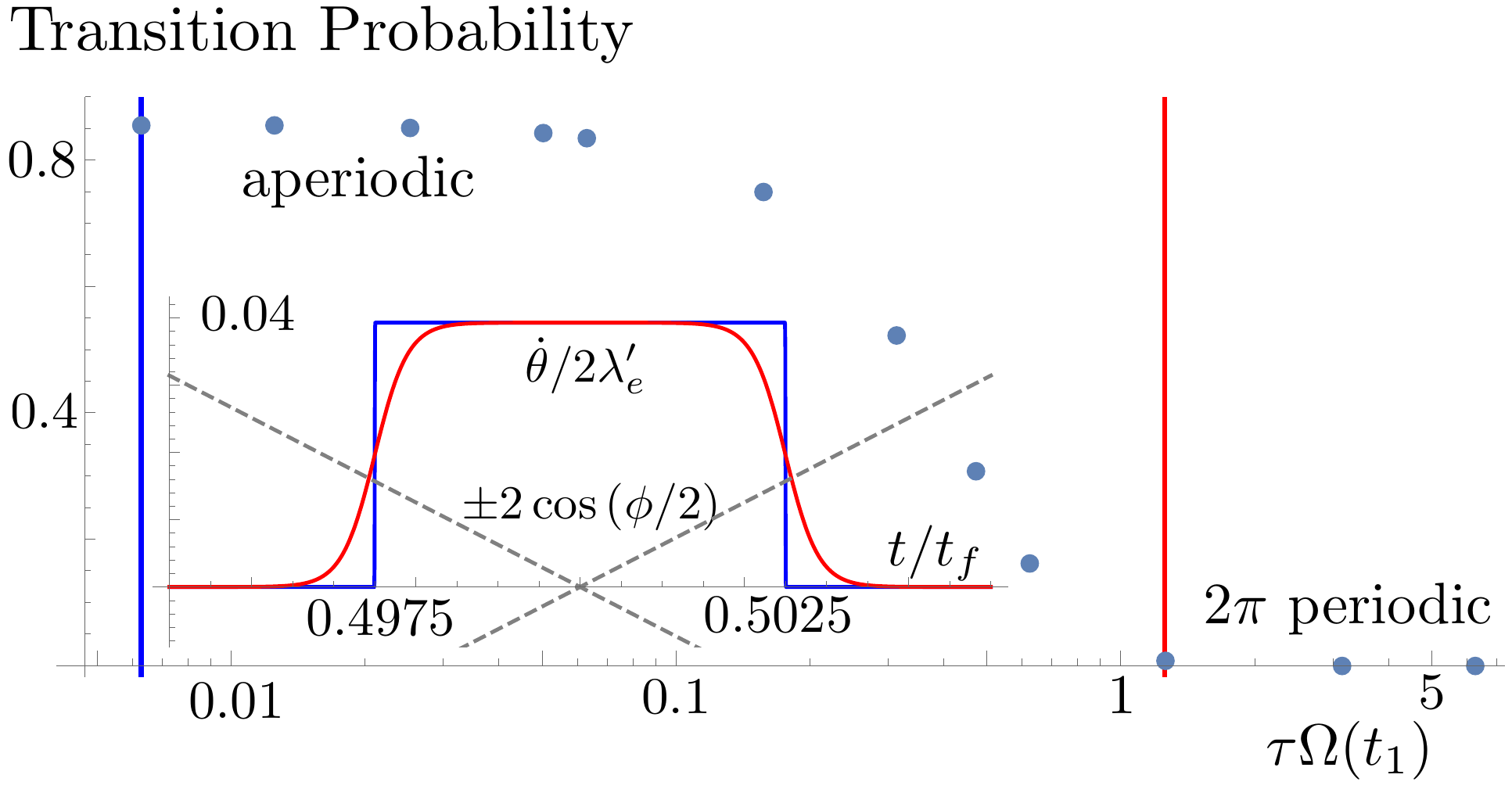}
	\caption{ Transition probability after evolution by $H_{\rm eff}(t)$ from $t=0$ to $t_f$ for $\alpha(t) = 2\pi n \left[ f(t)-f(t_i)\right]/\left[f(t_f)-f(t_i)\right]$ with ${f(t)= \log\left[ \left( 1+ e^{(t-t_1)/\tau}\right) / \left( 1+ e^{(t-t_2)/\tau} \right) \right]}$ and ${\phi(t)=2\pi t/t_f}$.
	Parameters are $\lambda_e=\varepsilon/80$, $\beta=\varepsilon/40$, $t_f=2\times 10^{7^{}} \varepsilon^{-1}$, $t_{1/2}=t_f/2 \pm 0.0025 t_f$, and $n=25\times 10^3$.
	The choice of $\alpha(t)$ yields step-like jumps in $\dot\theta(t)$ over a time scale $\tau$, as shown in the inset.
	In the main plot's horizontal axis, $\Omega(t_1)$ denotes the magnitude of the instantaneous energy at the time of the jumps.
	`Slow' jumps [$\tau \Omega(t_1) \gtrsim 1$] yield nearly zero transition probability, signifying a $2\pi$-periodic Josephson effect.
	`Fast' jumps [$\tau \Omega(t_1) \lesssim 1$], however, generate aperiodicity---even for arbitrarily slow $\phi(t)$.
	{\it Inset}: Off-diagonal element of $H_{\rm eff}(t)/{\lambda_e'}$ near $\phi = \pi$ for $\tau \Omega(t_1) \approx 1.257$ (red) and $\tau \Omega(t_1) \approx 0.00629$ (blue), along with diagonal elements (grey).
	}\label{fig:trans-prob}
\end{figure}

We take $\varepsilon$ to be the largest energy scale and, for simplicity, project onto the even-parity subspace ${\gamma_L \gamma_R \tilde{\gamma}_L\tilde{\gamma}_R\gamma_\varepsilon \gamma_{\varepsilon'}\tilde{\gamma}_\varepsilon \tilde{\gamma}_\varepsilon' =+1}$.
In this formulation, the Hamiltonian $H(t)= H_\text{JJ} + \delta H(t)$ supports two `low-energy' instantaneous eigenstates---denoted $\ket{\psi_1(t)}$ and $\ket{\psi_2(t)}$---separated from the next lowest instantaneous eigenstates by an excess energy $\sim \varepsilon$.
We further assume that $\alpha(t)$ varies slowly in time, i.e., $\beta \dot \alpha(t) \ll \varepsilon^{2}$, so that transitions between the low- and high-energy states are negligible.
Solutions to Schr\"odinger's equation then approximately take the form $\ket{\Phi(t)} = v_1(t) \ket{\psi_1(t)} + v_2(t) \ket{\psi_2(t)}$; the
coefficients satisfy the equation of motion ${i \partial_t {\vec v} = [H_{\rm inst}(t) + H_{\rm B}(t)]{\vec v} \equiv H_{\rm eff}(t) \vec v}$, with $H_{\rm inst}$ a diagonal matrix populated by the instantaneous energies and $H_{{\rm B},ij} = -\bra{\psi_i(t)}i \partial_t \ket{\psi_j(t)}$ a Berry-phase term.
Retaining terms up to $\mathcal{O}\left(\varepsilon^{-2}\right)$ (except an unimportant term proportional to identity), we explicitly find
\begin{align}\label{eq:Heff}
  H_{\rm eff}(t) &= 2\lambda_e' \cos[\phi(t)/2]\sigma_z - \frac{1 }{2}\dot{\theta}(t)\sigma_y,
\end{align}
where $\lambda_e' = \lambda_e\left[1-\beta^2/(2\varepsilon^2)\right]$ is a renormalized tunneling amplitude, $\dot{\theta}(t) =- \dot{\alpha}(t) \beta^2/\varepsilon^2$, and the Pauli matrices now refer to the basis of instantaneous eigenstates $\ket{\psi_j(t)}$.
When $\lambda_e' = 0$, the time-evolution operator ${U = e^{-i \int_0^T H_{\rm eff}(t)} \equiv e^{i \frac{\theta}{2} \sigma_y}}$ implements local mixing among the degenerate junction states with mixing angle ${\theta = \int_0^T dt \,\dot{\theta}(t)}$,
in agreement with Ref.~\onlinecite{Wolms14}~\footnote{Up to a factor of two error in their Eq.~(8a).} at $\beta \ll \varepsilon$.
More interestingly, with $\lambda_e' \neq 0$, the $\sigma_y$ term responsible for local mixing {\it effectively} couples the bound states related by time-reversal symmetry---even though the crossings in Fig.~\ref{fig:Tm1} are protected.

To analyze the Josephson effect described by Eq.~\eqref{eq:Heff}, we first consider $\dot{\phi}$ and $\dot{\theta}$ approximately constant.
Treating local mixing as a small perturbation away from the time-reversal-invariant point, we expand $H_\text{eff}(t)$ near $\phi=\pi$ to obtain a standard Landau-Zener Hamiltonian.
The transition probability between instantaneous $H_{\rm eff}(t)$ eigenstates monotonically increases as $x = \dot{\theta}^2/\lambda_e'\dot{\phi}$ decreases.
At $x\gg 1$, the adiabatic criterion is satisfied;
here a system initialized into the instantaneous ground state at $\phi=0$ evolves into the instantaneous ground state at $\phi=2\pi$, yielding a $2\pi$-periodic current-phase relation.
For $x\ll 1$ `fast' phase winding instead overwhelms local mixing, and a $4\pi$-periodic fractional Josephson effect emerges.

Next we examine a `quench' that more closely resembles the proof-of-concept picture considered earlier:
As sketched in the inset of Fig.~\ref{fig:trans-prob}, during an interval at which $\phi \approx \pi$, $\dot \theta(t)$ jumps from zero to a finite value over a time scale $\tau$ and then similarly decays back to zero.
Figure~\ref{fig:trans-prob}, main panel, depicts the numerically obtained transition probability as a function of $\tau \Omega(t_1)$, where $2\Omega(t_1)$ is the instantaneous energy gap evaluated at the jump; see caption for parameters.
For `large' $\tau$ a local-mixing-induced $2\pi$-periodic Josephson effect again arises (Supplemental Material derives a condition for adiabatic evolution in this case).
As $\tau$ decreases, however, the transition probability becomes appreciable and eventually plateaus---indicating an aperiodic current-phase relation.

Thus far we have focused on unitary time evolution.
We now note that continuously measuring the current can stabilize $4\pi$ periodicity through the quantum Zeno effect.
Although current eigenstates correspond to energy eigenstates, the current is most distinguishable when the energies are degenerate; hence measurement backaction competes against local mixing.
If the measurement projects onto a current eigenstate faster than the time scale of local mixing, the fractional Josephson effect survives (up to processes not considered here).

\begin{figure}
	\includegraphics[width=\columnwidth]{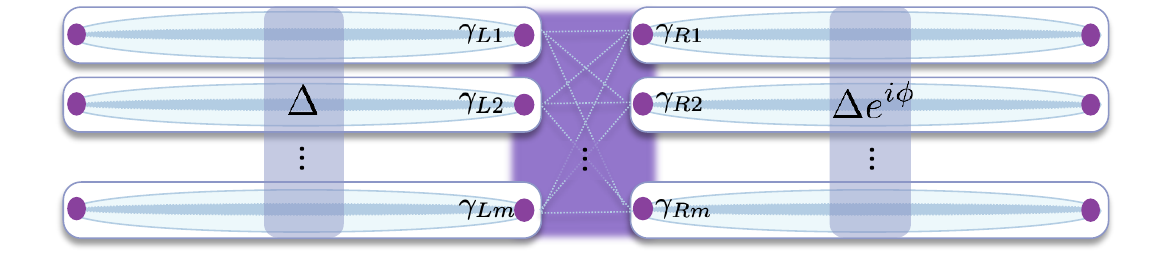}
\begin{tabular}{|c|c|c|}\hline
		~$m$ mod 4~ & non-interacting $I(\phi)$ & interacting $I(\phi)$  \\\hline
		~1~ & ~$4\pi$~ & ~$4\pi$~ \\ \hline
		~2~ & ~$2\pi$, $4\pi$, or aperiodic~ & $2\pi$, $4\pi$, or aperiodic~ \\  \hline
		~3~ & $4\pi$ or aperiodic~ & ~$4\pi$~ \\ \hline
		~4~ & ~$2\pi$, $4\pi$, or aperiodic~ & ~$2\pi$~  \\ \hline
\end{tabular}
	\caption{Josephson junction between two sets of $m$ proximitized nanowires that each (approximately) obey $\mathcal{T}^2 = +1$ time-reversal symmetry \cite{Tewari12}.
	{\it Table:} Summary of the periodicity for the current-phase relation in the presence of local mixing, both in the non-interacting limit and with interactions.  Multiple values are given when the result depends on local-mixing details.
}\label{fig:Tp1}
\end{figure}

%%%
{\bf \emph{$\mathcal{T}^2=+1$ Josephson junction.}}~Topological superconductors with $\mathcal{T}^2 = +1$ time-reversal symmetry can support an arbitrary number $m$ of MZMs at each end in the non-interacting limit, but only $m\mod 8$ with interactions \cite{Fidkowski09}.
As a physical realization, we envision an array of $m$ proximitized semiconductor nanowires in a magnetic field \cite{Lutchyn10,Oreg10,Lutchyn17,Vaitiekenas20}, for which the minimal low-energy Hamiltonian preserves $\mathcal{T}^2 = +1$ symmetry \cite{Tewari12}.
Figure~\ref{fig:Tp1}, top, sketches a Josephson junction assembled from such arrays.
Our goal is to explore the impact of local mixing and interactions on the Josephson effect.

\begin{figure}
	\includegraphics[width=\columnwidth]{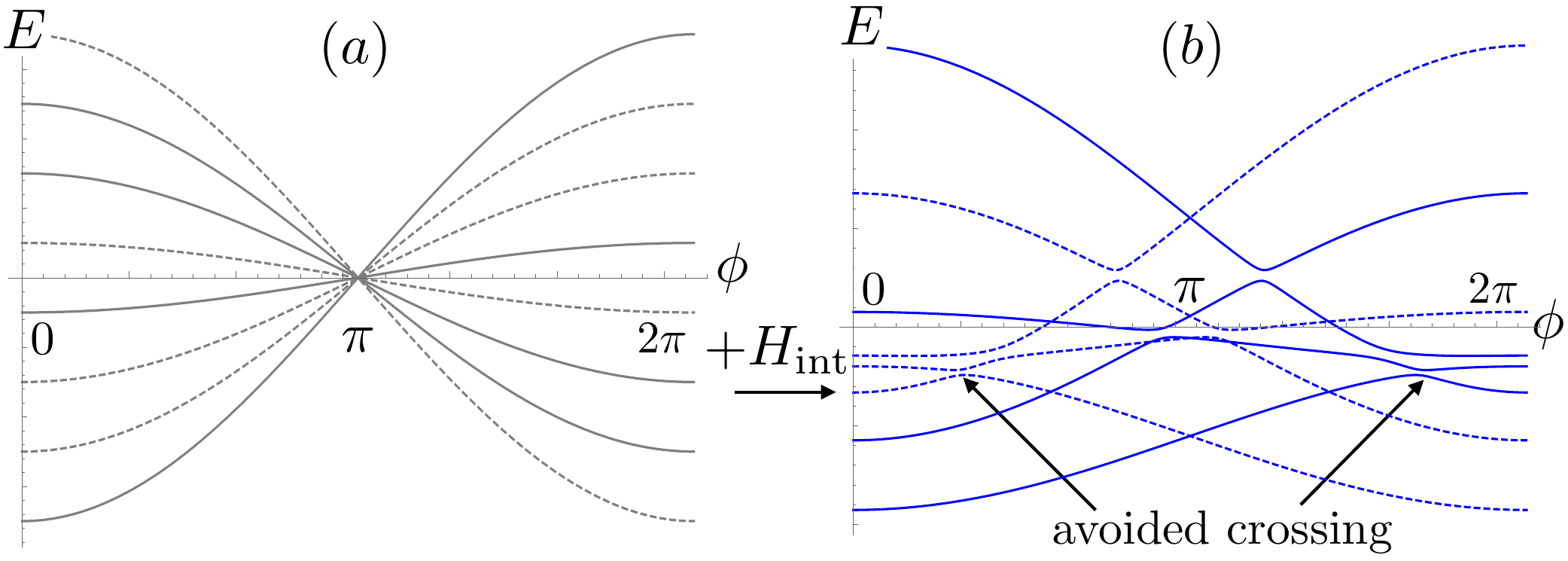}
	\caption{(a) Non-interacting many-body spectrum versus $\phi$ for the Josephson junction in Fig.~\ref{fig:Tp1} in the $m = 3$ case.
	Solid and dashed lines respectively denote even- and odd-fermion-parity states.
	Crossings necessarily occur at $\phi = \pi$ due to ${\cal T}^2 = +1$ symmetry, and local mixing generically rotates among equal-parity degenerate states.
	(b) Many-body spectrum with interactions---which shift the locations of the crossings away from the time-reversal-invariant point, where they are no longer protected.
	Interaction-induced avoided crossings between same-parity states protect the fractional Josephson effect against local mixing.
	}
	\label{fig:3wf}
\end{figure}

Consider first the non-interacting limit.
The junction hosts MZMs $\gamma_{L1,\ldots,m}$ from the left end and $\gamma_{R1,\ldots,m}$ from the right, which hybridize via
\begin{equation}
  H_{\rm JJ}^{\mathcal{T}^2 = +1} = \sum_{j,k} i\lambda_{jk} \cos(\phi/2) \gamma_{Lj}\gamma_{Rk}.
  \label{eq:HJJb}
\end{equation}
Time-reversal symmetry persists at $\phi = n\pi$ ($n \in \mathbb{Z}$); moreover, at $\phi = \pi$ the hybridization vanishes, yielding $2m$ MZMs at the junction.
As for the $\mathcal{T}^2 = -1$ case, each energy described by Eq.~\eqref{eq:HJJb} is $4\pi$ periodic in $\phi$, and hence a fractional Josephson effect exists at this level of analysis.
Figure~\ref{fig:3wf}(a) illustrates the energies versus $\phi$ for $m = 3$, with solid and dashed lines respectively denoting even- and odd-parity states.

To incorporate local mixing, observe that $\delta H(t)$ in Eq.~\eqref{eq:HLM} \emph{also} preserves $\mathcal{T}^2 = +1$ symmetry (with $\gamma_L,\tilde \gamma_L, \gamma'_{\varepsilon},\tilde \gamma'_{\varepsilon} \rightarrow \gamma_L,\tilde \gamma_L,\gamma'_{\varepsilon},\tilde \gamma'_{\varepsilon}$  and $\gamma_{\varepsilon},\tilde \gamma_{\varepsilon} \rightarrow -\gamma_{\varepsilon},-\tilde \gamma_{\varepsilon}$).
Thus one can immediately construct a local-mixing Hamiltonian for the $\mathcal{T}^2 = +1$ problem by replacing $\gamma_L \rightarrow \gamma_{Lj}, \tilde \gamma_L \rightarrow \gamma_{Lk}$ in $\delta H(t)$ and summing over $j,k$ pairs.
The net effect is that \emph{local mixing can once again nonuniversally rotate the system among same-parity Hamiltonian eigenstates that are degenerate at $\phi = \pi$}.
For any $m > 1$ this degeneracy is nontrivial, implying that local mixing spoils the fractional Josephson effect unless the phase is swept sufficiently rapidly.
Interestingly, for odd $m>1$ local mixing can never generate $2\pi$ periodicity since the junction parity switches upon sweeping $\phi$ by $2\pi$; see, e.g., Fig.~\ref{fig:3wf}(a).

Symmetry-preserving interactions, which we now turn on, substantially enrich this story.
First, one only needs to consider $m \mod 4$.  Indeed with $m = 4$ the junction at $\phi = \pi$ hosts 8 MZMs---whose degeneracy interactions completely obliterate \cite{Fidkowski09}, thus stabilizing $2\pi$ periodicity.
For $m \mod 4 = 1$ a given fermion-parity sector has a unique ground state at $\phi = \pi$, so the fractional Josephson effect is immune to local mixing.
The case $m \mod 4 = 2$ essentially reduces to the TRITOPS Josephson junction already examined in great detail; at $\phi = \pi$ a two-fold degeneracy in a given parity sector persists even with interactions, and local mixing accordingly generates $2\pi$-periodicity, $4\pi$-periodicity, or aperiodicity depending on details.
Finally, for $m \mod 4=3$ interactions shift crossings between same-parity states [recall Fig.~\ref{fig:3wf}(a)] away from $\phi = \pi$, where they become avoided crossings due to the absence of $\mathcal{T}$ symmetry; see Fig.~\ref{fig:3wf}(b).
Interactions consequently protect the fractional Josephson effect against local mixing.
See the Table from Fig.~\ref{fig:Tp1} for a summary.

%%%
{\bf \emph{Discussion.}}~A very general implication of our study is that symmetry-protected degeneracies among locally distinguishable states do not necessarily suffice for generating robust nontrivial adiabatic cycles; examination of dynamics under generic conditions is additionally required.
We have seen that $\mathcal{T}$-symmetric local mixing perturbations that render braiding non-universal in TRITOPS wires~\cite{Wolms14,Wolms16} also preclude a well-defined adiabatic fractional Josephson effect.
Similar results hold in junctions of $\mathcal{T}^2 = +1$ topological superconductors, with the interesting addition that interactions in some cases immunize against local mixing.
Our analysis exemplifies a more general result in Ref.~\onlinecite{McGinley20} that time-reversal-symmetry-protected effects are fragile in an open system.
Whether an analogous fate befalls cycles in systems with degeneracies protected by local unitary symmetries remains an interesting open question.

Our findings are relevant for experiments on both TRITOPS and nanowire-based Josephson junctions.
Recent experiments investigating the Josephson effect in proximitized quantum spin Hall edges~\cite{Bocquillon16,Deacon16,Bocquillon18} observed signatures of $4\pi$- and $2\pi$- periodicity, whereas theory predicts an $8\pi$-periodic Josephson effect~\cite{Zhang14,Orth15,Peng16,Hui17}.
Subgap energy levels corresponding to the same fermion parity are predicted to have a Kramers degeneracy at integer multiples of $\phi=\pi$; thus, local mixing could induce transitions at these time-reversal-invariant points.
Additionally, the degree of $\mathcal{T}^2=+1$ symmetry breaking in Majorana nanowires has important implications for topological quantum computing with MZMs~\cite{Karzig16,Plugge17,Vijay16}.
Projective MZM parity measurements proposed in Ref.~\onlinecite{Karzig16} rely on pairs of MZMs hybridizing through adjacent quantum dots.
Estimating the magnitude of symmetry breaking, e.g., by observing the time scale of $\dot{\phi}(t)$ for which the junction in Fig.~\ref{fig:Tp1} recovers a periodic Josephson effect, would bound the visibility of these measurements.

%%%

{\bf \emph{Acknowledgments.}}~We are grateful to Arbel Haim, Torsten Karzig, and Yang Peng for illuminating discussions.
We also thank Dima Pikulin and Charlie Marcus for conversations that stimulated this research.
This work was supported by the Army Research Office under Grant Award W911NF-17-1-0323;
the National Science Foundation through grant DMR-1723367;
the Caltech Institute for Quantum Information and Matter, an NSF Physics Frontiers Center with support of the Gordon and Betty Moore Foundation through Grant GBMF1250;
the Walter Burke Institute for Theoretical Physics at Caltech;
and the Gordon and Betty Moore Foundation’s EPiQS Initiative, Grant GBMF8682.

\appendix

\section{Local mixing review}\label{app:LM-review}

Reference~\onlinecite{Wolms14} derived that when the operators describing a Majorana Kramers pair depend on some parameters $\bm{ \eta}$, the local mixing angle is given by
\begin{align}\label{eq:theta-def}
 \theta &=\frac{1}{2} \oint d\bm{\eta} \{ \gamma(\bm{\eta}),\nabla_{\bm{\eta}}\tilde{\gamma}(\bm{\eta})\}.
\end{align}
We review a simple example of how a non-zero mixing angle can arise microscopically.

Consider a TRITOPS wire modeled by two Kitaev chains.
In the dimerized limit, the Hamiltonian is given by
\begin{align}\label{eq:H0}
H_0 &= \frac{\varepsilon}{2} \sum_{j,\sigma}i\gamma_{j b \sigma} \gamma_{j+1 a\sigma}
\end{align}
where $\sigma\in\{\uparrow,\downarrow\}$ labels the two time-reversed copies,  $a$ and $b$ the two Majorana flavors that form the spinful fermion, and $j$ the site.
The two Majorana operators corresponding to the same $j$ and $\sigma$ transform oppositely under $\mathcal{T}$: here we take the signs in Eq.~\eqref{eq:TRS-gamma} to be $s_a=-1,\, s_b=1$.
This model has four MZMs, $\gamma_{1a\sigma},\,  \gamma_{Nb\sigma}$.
Let us assume that local perturbations on the left end of the wire take the form of a chemical potential $H_\mu$ and $s$-wave pairing $H_\Delta$:
\begin{align}
H_\mu(t) &= \frac{\beta\cos\alpha(t)}{2} \sum_\sigma \left(i\gamma_{1a\sigma}\gamma_{1b\sigma}+1\right) \label{eq:Hmu}
\\ H_\Delta(t)&= \frac{\beta \sin\alpha(t)}{2} \left( i\gamma_{1a\uparrow}\gamma_{1b\downarrow} -i\gamma_{1a\downarrow}\gamma_{1b\uparrow}\right), \label{eq:HDelta}
\end{align}
where $\alpha$ parametrizes the ratio of the two terms and is time-dependent.
Both Eq.~\eqref{eq:Hmu} and Eq.~\eqref{eq:HDelta} commute with $\mathcal{T}$.
In the presence of these perturbations, the new zero mode operators become time-dependent as well:
\begin{align}
\gamma_1 (t)&= \cos\zeta \gamma_{1a\uparrow}-\sin\zeta \left(\cos\alpha(t) \gamma_{2a\uparrow}+\sin\alpha(t)\gamma_{2a\downarrow} \right)
\\ \tilde{\gamma}_1(t) &=  \cos\zeta \gamma_{1a\downarrow}-\sin\zeta \left(\cos\alpha(t) \gamma_{2a\downarrow}-\sin\alpha(t)\gamma_{2a\uparrow} \right),
\end{align}
where $\tan\zeta=\beta/\varepsilon$.
Solving Eq.~\eqref{eq:theta-def}, we have
\begin{align}\label{eq:mixing-ex}
\theta_1 &= -\sin^2\zeta \oint d\alpha = -\sin^2\zeta\int_0^T dt\, \dot{\alpha}(t),
\end{align}
where $\alpha(T)=\alpha(0)+2\pi n$, with $n\in \mathbb{Z}$.
Therefore, provided $\alpha$ has non-trivial winding, $\theta_1\neq 0$ and $\gamma_1$, $\tilde{\gamma}_1$ undergo local mixing.

\begin{figure}
	\includegraphics[width=\columnwidth]{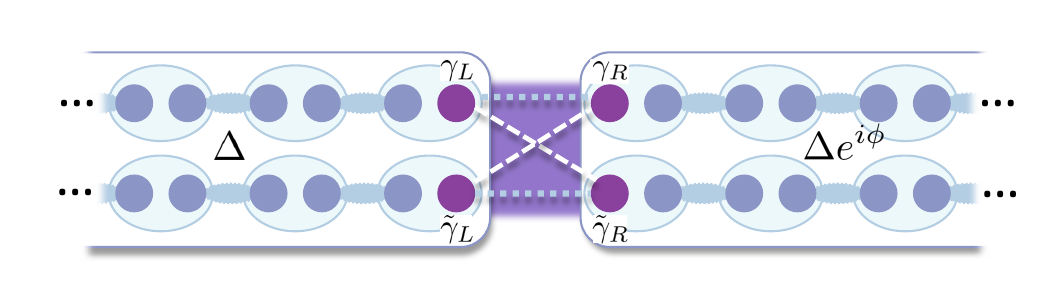}
	\caption{TRITOPS Josephson junction with each wire modeled by two Kitaev chains in the dimerized limit.
	Dotted lines indicate hybridization of Majorana Kramers pairs (dark purple) leading to Eq.~\eqref{eq:HJJ}.
	Local mixing arises when Majorana Kramers pairs undergo time-dependent coupling to gapped Majorana modes.
	}\label{fig:app}
\end{figure}

\section{TRITOPS Josephson junction}\label{app:JJ-deriv}

A TRITOPS wire can be thought of as two topological superconductors related by time reversal symmetry.
Labeling the two copies with a spin degree of freedom $\sigma\in \{\uparrow,\downarrow\}$, time reversal acts on the fermionic operators of the $J$th wire as $\bm{c}_J=(c_{J,\uparrow},c_{J,\downarrow})^T$ as~\cite{Haim18}
\begin{align}\label{eq:TcT}
\mathcal{T} \bm{c}_J \mathcal{T}^{-1} &=  i s_J(\phi) e^{i\phi_J}\sigma_y \bm{c}_J.
\end{align}
The sign $s_J(\phi)=\pm 1$ represents a $\mathbb{Z}_2$ gauge-freedom when defining symmetry transformations of superconductors.
When multiple TRITOPS are present but disconnected, each satisfies its own time reversal symmetry according to Eq.~\eqref{eq:TcT}.  When two TRITOPS are connected, e.g. by a Josephson junction, the global symmetry transformation must be consistent between the two.  Therefore, a TRITOPS Josephson junction is only symmetric under $\mathcal{T}$ when the phase difference between the left and right superconductors is a multiple of $\pi$.  We label these discrete values the ``time-reversal-invariant points" and fix $s_L(\phi_L)=+1$ and $s_R(\phi_R=n\pi+\phi_L)=-(-1)^n$ below.

A simple model of a TRITOPS Josephson junction is
\begin{align}\label{eq:HJJ-simple}
H_\text{JJ} &=  \tilde{\lambda} c_{L\uparrow}^\dagger c_{R\uparrow} - \tilde{\lambda}^* c^\dagger_{L\downarrow}c_{R\downarrow} +\lambda c_{L\uparrow}^\dagger c_{R\downarrow} +\lambda^*c^\dagger_{L\downarrow}c_{R\uparrow} + h.c.
\end{align}
where $L/R$ denote whether the fermion belongs to the wire on the left/right end of the junction and we can generically allow for different tunneling amplitudes between wires with the same and different $\sigma$ labels.

Each fermionic operator can be written as
\begin{align}\label{eq:cdef}
c_{J\sigma} &= \frac{e^{-\frac{i\phi_J}{2}}}{2} \left(\gamma_{J a\sigma}+i\gamma_{J b\sigma}\right),
\end{align}
where $\phi_J$ is the superconducting phase of the wire on the $J$th side of the junction ($J\in\{L,R\}$), and the operators $\gamma_{c\sigma}$, ${c\in\{a,b\}}$  satisfy the Majorana anticommutation relation
\begin{align}
\{\gamma_{c\sigma},\gamma_{c'\sigma'}\}=2\delta_{cc'}\delta_{\sigma\sigma'}.
\end{align}
Equations~\eqref{eq:TcT} and Eq.~\eqref{eq:cdef} imply Eq.~\eqref{eq:TRS-gamma}.

Each copy of a topological superconductor has a single MZM at its end point.
Projecting to the low-energy subspace takes
\begin{align}\label{eq:projection}
c_{L\sigma} &\to \frac{e^{-\frac{i\phi_L}{2}}}{2} \gamma_{La\sigma}, & c_{R\sigma} &\to \frac{ie^{-\frac{i\phi_R}{2}}}{2}  \gamma_{Rb\sigma}.
\end{align}
Fixing $\phi_L=0$ and $\phi_R=\phi$ and dropping the $a/b$ label of the Majorana operators, Eq.~\eqref{eq:HJJ} becomes
\begin{widetext}
\begin{align}\label{eq:HJJ-full}
H_\text{JJ} &= \frac{1}{2}\sum_{\sigma=\uparrow/\downarrow}\left[\cos(\phi/2)\left(\sigma \text{Re}[\tilde{\lambda}]\, i\gamma_{L\sigma}\gamma_{R\sigma}+ \text{Re}[{\lambda}] \, i\gamma_{L\sigma}\gamma_{R\bar{\sigma}}  \right) +\sin(\phi/2)\left(   \text{Im}[\tilde{\lambda}]\, i\gamma_{L\sigma}\gamma_{R\sigma}+\sigma \text{Im}[{\lambda}]\, i\gamma_{L\sigma}\gamma_{R\bar{\sigma}} \right)\right].
\end{align}
\end{widetext}
In the above, we have written $\bar{\sigma}$ to indicate the opposite choice of $\sigma$ for the subscripts, and $\sigma$ as a coefficient to correspond to $\pm$ for $\uparrow/\downarrow$.
We recover Eq.~\eqref{eq:HJJ} by setting $\tilde{\lambda}=0$ for simplicity, denoting the real and imaginary parts of $\lambda$ with subscripts $e/o$, and identifying ${(\gamma_{L,\uparrow},\gamma_{L,\downarrow},\gamma_{R,\uparrow},\gamma_{R,\downarrow})\to (\gamma_L,\tilde{\gamma}_L,\tilde{\gamma}_R,\gamma_R)}$.

\subsection{Deriving $H_\text{eff}$}

We now derive Eq.~\eqref{eq:Heff} in the main text.  The Majorana operators in Eqs.~\eqref{eq:HJJ} and \eqref{eq:HLM} can be written in terms of complex fermionic operators as
\begin{align}
f^\dagger &= \frac{1}{2} \left( \gamma_\varepsilon' - i\gamma_{\varepsilon} \right)
\\ \tilde{f}^\dagger &= \frac{1}{2} \left(\tilde{\gamma}'_\varepsilon - i \tilde{\gamma}_\varepsilon \right)
\\ c^\dagger &= \frac{1}{2} \left( \gamma_L + i\gamma_R \right)
\\ \tilde{c}^\dagger &= \frac{1}{2} \left( \tilde{\gamma}_L + i\tilde{\gamma}_R \right)
\end{align}
Then, defining even-parity basis states so that $\ket{0}$ corresponds to the vacuum state annihilated by $c, \tilde{c}, f, \tilde{f}$ and
\begin{align}
 \ket{1} &= c^\dagger \tilde{c}^\dagger \ket{0}
 \\ \ket{2} &= f^\dagger c^\dagger \ket{0}
 \\ \ket{3} &= f^\dagger \tilde{c}^\dagger \ket{0}
 \\ \ket{4} &= \tilde{f}^\dagger c^\dagger \ket{0}
 \\ \ket{5} &= \tilde{f}^\dagger \tilde{c}^\dagger \ket{0}
 \\ \ket{6} &= f^\dagger \tilde{f}^\dagger \ket{0}
 \\ \ket{7} &= f^\dagger \tilde{f}^\dagger c^\dagger \tilde{c}^\dagger \ket{0}
\end{align}
so that  $f^\dagger f = \frac{1}{2} \left( 1-i\gamma_\varepsilon \gamma'_\varepsilon \right)$ and $c^\dagger c = \frac{1}{2} \left( 1- i\gamma_L\gamma_R\right)$ and similarly for the time-reversed partners.
In this basis, the full Hamiltonian can be written in first-quantized form as
\begin{widetext}
\begin{align}\label{eq:HKit-full}
H &= \left( \begin{array}{cc|cccc|cc}
\bm{\lambda_e} & 0 & \frac{\beta}{2}\cos\alpha & -\frac{\beta}{2}\sin\alpha & \frac{\beta}{2}\sin\alpha & \frac{\beta}{2} \cos\alpha & 0 & 0
\\ 0 &- \bm{\lambda_e} & \frac{\beta}{2} \sin\alpha & \frac{\beta}{2}\cos\alpha & -\frac{\beta}{2} \cos\alpha & \frac{\beta}{2}\sin\alpha & 0 & 0
\\ \hline
\frac{\beta}{2}\cos\alpha & \frac{\beta}{2}\sin\alpha & \varepsilon - \bm{\lambda_o} & 0 & 0 & 0 & \frac{\beta}{2}\sin\alpha & -\frac{\beta}{2} \cos \alpha
\\ -\frac{\beta}{2} \sin\alpha & \frac{\beta}{2}\cos\alpha & 0 & \varepsilon + \bm{\lambda_o} & 0 & 0 & \frac{\beta}{2}\cos\alpha & \frac{\beta}{2}\sin\alpha
\\ \frac{\beta}{2}\sin\alpha & -\frac{\beta}{2}\cos\alpha & 0 & 0 & \varepsilon - \bm{\lambda_o} & 0 & -\frac{\beta}{2}\cos\alpha & -\frac{\beta}{2}\sin\alpha
\\ \frac{\beta}{2}\cos\alpha & \frac{\beta}{2} \sin\alpha & 0 & 0 & 0 & \varepsilon + \bm{\lambda_o} & \frac{\beta}{2}\sin\alpha & - \frac{\beta}{2}\cos\alpha
\\ \hline
0 & 0 & \frac{\beta}{2}\sin\alpha & \frac{\beta}{2} \cos\alpha & - \frac{\beta}{2} \cos\alpha & \frac{\beta}{2} \sin\alpha & 2\varepsilon + \bm{\lambda_e} & 0
\\ 0 & 0 & -\frac{\beta}{2}\cos\alpha & \frac{\beta}{2}\sin\alpha & -\frac{\beta}{2}\sin\alpha & - \frac{\beta}{2}\cos\alpha & 0 & 2\varepsilon - \bm{\lambda_e}
\end{array}\right)
\end{align}
\end{widetext}
where we have adopted the shorthand $\bm{\lambda_e}=2\lambda_e \cos \left(\frac{\phi}{2}\right)$ and $\bm{\lambda_o}=2\lambda_o\sin\left(\frac{\phi}{2}\right)$ and suppressed the time-dependence of $\phi$ and $\alpha$.
Working to order $\varepsilon^{-2}$, the two lowest energies are
\begin{align}
\epsilon_{1/2}(t) &= \pm 2\lambda_e\left( 1- \frac{\beta^2}{2\varepsilon^2}\right) \cos\left( \frac{\phi(t)}{2}\right)
\end{align}
with corresponding instantaneous eigenstates
\begin{widetext}
\begin{align}
\ket{\psi_1(t)} &=+ \ket{1} - \frac{\beta}{2\varepsilon} \left( \sin \alpha(t) \left[ \nu_{-+}(t)\ket{2}+\nu_{--}(t)\ket{5}\right] -\cos\alpha(t)\left[\nu_{--}(t)\ket{3} - \nu_{-+}(t)\ket{4}\right] \right) - \left( \frac{\beta}{2\varepsilon}\right)^2 \left( \ket{1} -\ket{6} \right).
%%%
\\ \ket{\psi_2(t)} &=
 -\ket{0} + \frac{\beta}{2\varepsilon}\left( \cos \alpha(t) \left[ \nu_{++}(t)\ket{2}+\nu_{+-}(t)\ket{5}\right] -\sin\alpha(t)\left[\nu_{+-}(t)\ket{3} - \nu_{++}(t)\ket{4}\right] \right)  + \left( \frac{\beta}{2\varepsilon}\right)^2 \left( \ket{0}+ \ket{7} \right)
\end{align}
\end{widetext}
We have defined $\nu_{p,p'} = 1 + \left(-p {\bm \lambda_e}(t) + p'{\bm \lambda_o}(t)\right)/\varepsilon$ with $p,p'=\pm 1$.

As described in the main text, when $ \beta \dot{\alpha}(t) \ll \varepsilon^{2}$, transitions between the low and high-energy states are negligible.  Solutions to the Schr\"odinger equation for a state initialized in the low-energy subspace take the form
\begin{align}
\ket{\Phi(t)} &= v_1(t) \ket{\psi_1(t)} + v_2(t)\ket{\psi_2(t)}.
\end{align}
The coefficients satisfy the equation of motion
\begin{align}
i \partial_t \bm{v} &=\left[ H_\text{inst}(t) + H_\text{B}(t)\right]\bm{v} = H_\text{eff}(t) \bm{v}
\end{align}
where
\begin{align}
H_\text{inst}(t) &= 2\lambda_e \left( 1 - \frac{\beta^2}{2\varepsilon^2} \right) \cos\left( \frac{\phi(t)}{2} \right) \sigma_z
\\ H_\text{B}(t) &= \bra{\psi_1(t)}\partial_t \ket{\psi_2(t)} \sigma_y =  \frac{\dot{\alpha}}{2} \frac{\beta^2}{\varepsilon^2} \sigma_y,
\end{align}
recovering Eq.~\eqref{eq:Heff}:
\begin{align}
H_\text{eff}(t) &= 2\lambda_e' \cos \left[\phi(t)/2\right]\sigma_z + \dot{\alpha}(t) \frac{\beta^2}{2\varepsilon^2} \sigma_y.
\end{align}
Note that all $\lambda_o$ dependence drops out at order $\varepsilon^{-2}$.

\subsection{Time evolution according to $H_\text{eff}(t)$}

Equation~\eqref{eq:Heff} has the general form $H(t)=a(t)\sigma_z+b(t)\sigma_y$ with instantaneous eigenvalues and eigenstates
\begin{align}
\epsilon_\pm(t) &= \pm \Omega(t) = \pm \sqrt{a(t)^2+b(t)^2}
\\ \ket{\pm(t)} &= \mp i \beta_\pm \ket{0} + \beta_\mp \ket{1}
\end{align}
where $\ket{0}$, $\ket{1}$ are the eigenstates of $\sigma_z$ corresponding to eigenvalues $\pm1$, respectively, and we have defined ${\beta_\pm(t) = \sqrt{\frac{\Omega(t)\pm a(t)}{2\Omega(t)}}}$.
Consider the parameter
\begin{align}
A &= \frac{\text{max}_t |\bra{+(t)}\dot{H}_\text{eff}(t) \ket{-(t)}|}{4\text{min}_{t'} \, \Omega(t')^2}
\\ &= \frac{\text{max}_t \left[ |-\dot{a}(t) b(t) + \dot{b}(t) a(t)|/\Omega(t)\right]}{4\text{min}_{t'} \,\Omega(t')^2}.
\end{align}
The adiabatic theorem asserts that when $A\ll 1$, the system initialized in an energy eigenstate remains in that eigenstate throughout the time evolution~\cite{Born28,Lidar09}.

For $H_\text{eff}(t)$, $A$ and $\Omega(t)$ evaluate to
\begin{align}
A &=  \frac{\lambda_e'|- \cos[\phi(\tilde{t})/2]\ddot{\theta}(\tilde{t})-\dot{\phi}(\tilde{t}) \sin[\phi(\tilde{t})/2] \dot{\theta}(\tilde{t})/2 |/\Omega(\tilde{t})}{4 \Omega(\bar{t})^2}
\\ \Omega(t) &= \sqrt{4\lambda_e'^2\cos^2[\phi(t)/2] + \dot{\theta}(t)^2/4},
\end{align}
where $\tilde{t}$ is the time that maximizes the numerator and $\bar{t}$ the time that minimizes the denominator.
When $\ddot{\theta}(t)=0$, $A$ reduces to the inverse of the Landau-Zener parameter $x$:
\begin{align}
A^\text{LZ} &= \frac{\lambda_e' \dot{\phi}(t^*)}{\dot{\theta}^2} =\frac{1}{x}.
\end{align}
Alternatively, for the quench considered in Fig.~\ref{fig:trans-prob}, $\dot{\phi}\to 0$ and $A$ becomes
\begin{align}
A^\text{quench} &= \frac{\lambda_e' \cos[\phi(t_1)/2] \ddot{\theta}(t_1)/\Omega(t_1)}{4 \Omega(\bar{t})^2},
\end{align}
where $t_1$ is the location of the quench.
If ${\dot{\theta}_\text{max}\ll 2\lambda_e'\cos[\phi(t_1)/2]}$, then the denominator reduces to $\dot{\theta}_\text{max}^2$ and
\begin{align}
A^\text{quench} &\approx \frac{\ddot{\theta}(t_1)}{{2} \dot{\theta}_\text{max}^2} \approx \frac{1}{{2} \tau \dot{\theta}_\text{max}}.
\end{align}
Thus the transition probability approaches zero for ${\tau \dot{\theta}_\text{max} \gg 1/2}$.
If instead $\dot{\theta}_\text{max}/2$ and $2\lambda_e'\cos[\phi(t_1)/2]$ are comparable (as is the case in Fig.~\ref{fig:trans-prob}), the adiabatic criterion becomes $8 \tau \Omega(t_1) \gg 1$.

To analyze the time evolution according to Eq.~\eqref{eq:Heff} more generally, we can consider the Schr\"odinger equation for a state ${\ket{\psi(t)}=\sum_{\sigma=\pm} c_\sigma(t)\ket{\sigma(t)}}$.
The  coefficients $c_\pm(t)$ satisfy
\begin{align}
i \left( \begin{array}{c} \dot{c}_+(t)  \\ \dot{c}_-(t) \end{array}\right) &=\left(  \Omega(t)\sigma_z +  v(t) \sigma_y \right) \left( \begin{array}{c} c_+(t) \\ c_-(t) \end{array}\right)
\end{align}
where
\begin{align}
v(t)& \equiv \bra{+(t)}\partial_t \ket{-(t)}
= \frac{\dot{a}(t) b(t) - \dot{b}(t) a(t)}{2\Omega(t)^2}
\\ &= \frac{\lambda_e' \left( \dot{\phi} \sin[\phi/2] \dot{\theta}/2 + \cos[\phi/2] \ddot{\theta}\right)}{2\Omega^2}.
\end{align}
When $|v(t)|\ll \Omega(t)$, the coefficients evolve according to a diagonal Hamiltonian and the system initialized in the instantaneous ground state will remain in the instantaneous ground state at later times, resulting in the conventional $2\pi$-periodic Josephson effect.
(Note that $\text{max}_t |v(t)|/\Omega(t)$ corresponds to $A$ when $\tilde{t}=\bar{t}$.)
When $|v(t)|\gg \Omega(t)$, the instantaneous energy states undergo Rabi oscillations, and the current-phase relation will generally be aperiodic.

%%%
\section{Aperiodicity from local mixing}

Consider a junction described by Eq.~\eqref{eq:HJJ-full}.
In the even parity sector ${i\gamma_L \gamma_R=i\tilde{\gamma}_L\tilde{\gamma}_R}$, we can define Pauli matrices
\begin{align}
 \sigma^x &= i\gamma_L \gamma_R = i\tilde{\gamma}_L \tilde{\gamma}_R
\\ \sigma^y &= i\gamma_L\tilde{\gamma}_R =- i\tilde{\gamma}_L\gamma_R
\\ \sigma^z &= i\gamma_L \tilde{\gamma}_L = i\gamma_R \tilde{\gamma}_R
\end{align}
so that
\begin{align}
H_\text{JJ}^{(e)}(t) &= 2\sqrt{\lambda_e^2+\tilde{\lambda}_e^2} \cos\left(\frac{\phi(t)}{2}\right) \left(\begin{array}{cc} 0 & a-ib \\ a+ib & 0 \end{array}\right)
\end{align}
for  $a=\lambda_e/\sqrt{\lambda_e^2+\tilde{\lambda}_e^2},$ $b=\tilde{\lambda}_e/\sqrt{\lambda_e^2+\tilde{\lambda}_e^2}$.  When $\phi$ is not equal to an odd multiple of $\pi$, the junction eigenstates are
\begin{align}
\ket{I_\pm} &= \frac{1}{\sqrt{2}} \left( \begin{array}{c} a-ib \\ \pm 1\end{array}\right).
\end{align}
Note that $\ket{I_-}$ is the instantaneous ground state of the junction for $\phi<\pi$, while $\ket{I_+}$ is the instantaneous ground state for $\phi>\pi$.
Consider a thought experiment of a phase-biased TRITOPS Josephson junction, undergoing the following protocol.
Initialize the system at $\phi=0$ in the state $\ket{I_+}$, then evolve the phase $\phi$ such that at the $k$th time invariant point one of the Majorana Kramers pairs accrues a local mixing angle $\theta_k$.
In the absence of any other noise sources, between the $k-1$th and $k$th time-reversal invariant points, the system is in a superposition of junction eigenstates
\begin{align}
\ket{\psi_{12}} &= \cos\left(\frac{\sum_k\theta_k}{2}\right) \ket{I_+} + i\sin\left(\frac{\sum_k\theta_k}{2}\right) \ket{I_-}
\end{align}
with current expectation value
\begin{align}
\langle I (\phi) \rangle &=- \frac{e}{\hbar}\sqrt{\lambda_e^2+\tilde{\lambda}_e^2} \cos \left(\sum_{j=1}^k \theta_j \right) \sin\left(\frac{\phi}{2}\right).
\end{align}
When $\theta_k\neq 2\pi$,
\begin{align}
\cos\left(\sum_{j=1}^{k-1} \theta_j\right) &\neq \cos\left(\sum_{j=1}^{k+1} \theta_j \right)
\end{align}
thus the current expectation value is not $4\pi$ periodic.
More generally, $\langle I(\phi)\rangle$ is aperiodic except for fine-tuned choices of the $\theta_j$.

The phase-biased system is not necessarily the most experimentally accessible, as usually phase would be tuned by a magnetic field, whose presence would break the time reversal symmetry of the junction.
A more physically relevant setup is for the junction to be voltage-biased, so that the DC Josephson equation implies a constant phase sweep speed $\dot{\phi}=2e V/\hbar=\omega_J$.
When the system undergoes a $4\pi$ periodic fractional Josephson effect, the power spectrum of the current
\begin{align}
P(\omega) &= \lim_{C\to \infty} \int_{0}^C dt \int_0^C dt' \langle I(t')I(t) \rangle e^{i\omega(t'-t)}
\end{align}
exhibits a peak at $\omega=\pm \omega_J/2$.
An aperiodic current-phase relation manifests as no peak in the power spectrum.

If the only source of noise is local mixing, then the probability $q_\pm$ of occupying junction eigenstates $\ket{I_\pm}$ only changes after passing through a time reversal invariant point.
If ${s_k=1-p_k}$ is the probability of transitioning between junction eigenstates (i.e. $p_k$ is the probability of transitioning between instantaneous energy eigenstates) at the $k$th such point, and $q_\pm(t_k)$ is the occupation probability of $\ket{I_\pm}$ preceding that point, then
\begin{align}
\left( \begin{array}{c} q_{+}(t_{k+1}) \\ q_{-}(t_{k+1}) \end{array}\right) &= \left( \begin{array}{cc} 1-s_k & s_k \\ s_k & 1-s_k \end{array}\right)  \left( \begin{array}{c} q_{+}(t_k) \\ q_{-}(t_k) \end{array}\right).
\end{align}
Approximating ${s_k=\sin^2\left(\theta_k/2\right)}$ by its average value, $\bar{s}$
\begin{widetext}
\begin{align}\label{eq:Sk}
\left( \begin{array}{c} q_{+}(t_{k+1}) \\ q_{-}(t_{k+1}) \end{array}\right)  &= \frac{1}{2} \left( \left[ 1+(1-2\bar{s})^k\right]\openone + \left[1-(1-2\bar{s})^k \right]\sigma_x \right) \left( \begin{array}{c} q_+(t_0) \\ q_-(t_0)\end{array}\right).
\end{align}
The matrix in Eq.~\eqref{eq:Sk} defines the propagator from $t_j$ to $t_{j+k}$:
\begin{align}
U\left(t=\frac{2\pi k}{\omega_J}\right) &=  \frac{1}{2} \left( \left[ 1+(1-2\bar{s})^k\right]\openone + \left[1-(1-2\bar{s})^k \right]\sigma_x \right).
\end{align}
Note that in the large $k$ limit the system approaches the maximally mixed state at a rate $\omega_J\ln[1-2\bar{s}]/2\pi$.

The current is
\begin{align}
I_\pm(t) &= \pm \frac{e}{\hbar}\sqrt{\lambda_e^2+\tilde{\lambda}_e^2} \sin\left(\frac{\omega_J}{2}t\right),
\end{align}
corresponding to correlator for $t'>t$~\cite{Pikulin11}
\begin{align}
\langle I(t')I(t)\rangle &= \sum_{ij=\pm} I_i(t')I_j(t) U_{ij}(t'-t)
 = 2I_0^2 \sin\left(\frac{\omega_J t}{2} \right)\sin\left(\frac{\omega_J t'}{2}\right) \left(1-2 \bar{s}\right)^{\frac{\omega_J (t'-t)}{2\pi} }
\\ &= 2 I_0^2 \sin \left( \frac{\omega_J t}{2}\right)\sin\left(\frac{\omega_J t'}{2} \right) \left( e^{\frac{\omega_J}{2\pi} \ln \left[ 1-2\bar{s}\right](t'-t)}\Theta\left(1- 2\bar{s}\right) + e^{\frac{\omega_J}{2\pi} \left( i\pi + \ln \left[ 2\bar{s}-1\right]\right) (t'-t)}\Theta\left(2\bar{s}-1\right) \right)
\end{align}
for $I_0=\frac{e}{\hbar}\sqrt{\lambda_e^2+\tilde{\lambda}_e^2}$.
Therefore, the power spectrum is
\begin{align}
P(\omega) &= \lim_{C\to \infty} \frac{ 2I_0^2 }{C} \int_0^C dt \int_0^C dt' e^{i\omega(t'-t)}\sin\left(\frac{\omega_J t'}{2}\right)  \sin\left(\frac{\omega_Jt}{2}\right)
 \notag
 \\ &\quad \quad \quad  \quad \quad \quad \quad \quad \quad \quad   \times  \left( e^{\frac{\omega_J}{2\pi} \ln[1-2\bar{s}]|t-t'|} \Theta(1-2\bar{s}) + e^{i\frac{\omega_J}{2} (t-t')} e^{\frac{\omega_J}{2\pi} \ln[2\bar{s}-1]|t-t'|} \Theta(2\bar{s}-1)\right)
\\ &= \frac{I_0^2}{2 \pi \omega_J } \sum_{a=\pm 1} \left(  \frac{ \ln [1-2\bar{s}]}{\left(\frac{\omega}{\omega_J}+ \frac{a}{2}\right)^2 + \left(\frac{\ln[1-2\bar{s}]}{2}\right)^2} \Theta(1-2\bar{s}) + \frac{ \ln [2\bar{s}-1 ]}{\left(\frac{\omega}{\omega_J}+\frac{a-1}{2} \right)^2 + \left(\frac{\ln[2\bar{s}-1]}{2}\right)^2} \Theta(2\bar{s}-1)\right).
\end{align}
\end{widetext}
As ${\bar{s}\to 0}$ ($\bar{p}\to 1$, $r\gg 1$), the power spectrum has two peaks at ${\omega=\pm \omega_J/2}$, corresponding to a fractional Josephson effect.  When ${\bar{s}\to 1}$ ($\bar{p}\to 0$, $r\ll1$), the power spectrum peaks at $\omega=0,\,\omega_J$, corresponding to the standard $2\pi$-periodic Josephson effect.  As $\bar{s}\to 1/2$ from either side, $P(\omega)$ flattens- signaling an aperiodic current-phase relation.

%%%
\section{$\mathcal{T}^2=+1$ Josephson junctions}

Consider the model for a topological superconductor suggested by Refs.~\onlinecite{Lutchyn10,Oreg10}
\begin{align}
H = \int_x \psi^\dagger (-\frac{\partial_x^2}{2m}-\mu-h\sigma^x-i\alpha\sigma^y\partial_x) \psi + \Delta \psi_\uparrow\psi_\downarrow + H.c.
\end{align}
where spin indices have been suppressed, $h$ is a Zeeman term, $\alpha$ is the spin-orbit coupling, and $\Delta$ is the superconduting gap.
This Hamiltonian is symmetric under $\mathcal{T}=\mathcal{K}$ time-reversal-symmetry~\cite{Fidkowski09,Tewari12}, which in this model is simply complex conjugation.
This symmetry is an artifact of the low-energy Hamiltonian and can be broken by adding higher-order hopping terms or interactions.
Nonetheless, such terms are expected to be weak and for low energies the wire satisfies $\mathcal{T}^2 = +1$.

We now derive the Josephson junction Hamiltonian for the setup shown in Fig.~\ref{fig:Tp1} when each Majorana nanowire individually satisfies $\mathcal{T}$.
Label the fermionic operators by $c_{Jj}$, $J\in\{L,R\}$ labeling the left/right side of the junction, and $j\in\{1,2\}$ labeling the top or bottom wire.
The $c_{J,j}$ transform trivially under $\mathcal{T}$; thus the most general non-interacting Hamiltonian describing the Josephson junction that is even under $\mathcal{T}$ is
\begin{align}\label{eq:HJJ1}
H_\text{JJ}^{(+1)} &= \sum_{J,j\neq k} \Lambda_{Jjk} c_{Jj}^\dagger c_{Jk} + \sum_{j,k} \left( 2\lambda_{jk} c_{Lj}^\dagger c_{Rk} + h.c\right).
\end{align}
where all tunneling amplitudes are real: $\Lambda_{Jjk}$, $\lambda_{jk} \in \mathbb{R}$.

Time reversal symmetry acts on the complex fermionic operators $c_{Jj}= \frac{e^{-i\phi_J/2}}{2} \left( \gamma_{Jaj} + i\gamma_{Jbj}\right)$ as $c_{Jj}\to s_J(\phi_J)e^{i\phi_J}c_{Jj}$.  Thus, we once again see that $\phi=\phi_R-\phi_L=n\pi$ are the time-reversal invariant points.
Fixing $\phi_L=0$ and $\phi_R=\phi$, the transformation on the Majorana operators is
\begin{align}
\gamma_{Jaj} &\to s_J \gamma_{Jaj}, \quad \gamma_{Jbj} \to -s_J \gamma_{Jbj}
\end{align}
with signs $s_L=1$, $s_R=(-1)^n$ for $\phi=n\pi$.

Projection to the low-energy subspace takes the same form as Eq.~\eqref{eq:projection}
\begin{align}\label{eq:projection2}
c_{Lj} &\to \frac{e^{-\frac{i\phi_L}{2}}}{2} \gamma_{Laj} & c_{Rj} &\to \frac{ie^{-\frac{i\phi_R}{2}}}{2}  \gamma_{Rbj}.
\end{align}
From here on, we drop the $a/b$ labels and write the zero mode operators as $\gamma_{Jj}$.
Under $\mathcal{T}$,
\begin{align}
i\gamma_{J1}\gamma_{J2}& \to -i\gamma_{J1}\gamma_{J2} \label{eq:same-side}
\\ i\gamma_{Lj}\gamma_{Rk} &\to s_L s_R  i \gamma_{Lj} \gamma_{Rk}=(-1)^n i \gamma_{Lj}\gamma_{Rk}.
\end{align}
Equation~\eqref{eq:same-side} implies $\Lambda_J=0$ (and is precisely why in the presence of $\mathcal{T}$ the quantum dot-based MZM parity measurement proposed in Ref.~\onlinecite{Karzig16} does not work).
Therefore, we recover Eq.~\eqref{eq:HJJb}
\begin{align}
H_\text{JJ}^{(+1)} &=\sum_j i \lambda_{jk}  \cos\left(\frac{\phi}{2}\right) \gamma_{Lj}\gamma_{Rk} .
\end{align}

The model given in Eq.~\eqref{eq:HKit-full} is purely real and thus also satisfies $\mathcal{T}^2=+1$ symmetry.
As such, the derivation of Eq.~\eqref{eq:Heff} similarly holds for this system as well.

\subsection{Multiwire topological Josephson junctions}

We investigate the effect of local mixing on a Josephson junction between two sets of $m$ Majorana wires.
Above we argued that $m=2$ reproduces the aperiodic behavior of a TRITOPS junction.
We now demonstrate that interactions restore $4\pi$ periodicity for $m=3$, and $2\pi$ periodicity for $m=4$.
Such Josephson junctions offer a testbed for probing the $\mathbb{Z}_8$ classification of Majorana nanowires theorized by Ref.~\onlinecite{Fidkowski09}.

We consider the low-energy Hamiltonian
\begin{align}
H = \sum_j E_j \cos\left({\frac{\phi}{2}} \right)\, i\gamma_{Lj}\gamma_{Rj},
\end{align}
where $j$ runs over each of the $m$ wires and $L$ and $R$ signify the wires to the left and right of the junction.
After one evolution $\gamma_{Rj} \rightarrow -\gamma_{Rj}$.
We can combine Majorana fermions into Dirac fermions as $c_j = \gamma_{Lj} + i\gamma_{Rj}$.
After one evolution the occupation of this bound state switches.
Notice that for $m$ wires we track $2^m$ bound states, which for free fermions all intersect at $\phi = \pi$ (where all energies are $0$).
\begin{itemize}
\item $m=1$.
The standard fractional Josephson is immune to local mixing, as the two bound states differ by local fermion parity.
No local mixing terms are allowed that mix the states at $\phi = \pi$.

\item $m=2$.
The model posited in previous Appendices still respects $\mathcal{T}^2 = +1$ symmetry.
The four states in question split into even and odd parity states.
Unlike the $m = 1$ wire, however, fermion parity in the junction remains the same after a $2\pi$ evolution (as both bound states switch occupation) and so we can restrict ourselves to the even parity sector.
The crossing at $\phi = \pi$ is protected by our symmetry, but that does not prevent local mixing.

Interactions do not play an important role for $m = 2$.
The only acceptable interaction at $\phi = \pi$ reads
\begin{align}
H_\text{int} = w_1 (i\gamma_{L1}\gamma_{R1})(i\gamma_{L2}\gamma_{R2}),
\end{align} which only splits the even and odd parity sectors and does not affect the Josephson periodicity.
We recover local mixing, implying (for certain parameter regimes) the loss of $4\pi$ periodicity.

\item $m=3$.
While it may seem that $m = 3$ wires will suffer from local mixing as well, interactions conspire to restore $4\pi$ periodicity (in much the same way that interactions stabilize an $8\pi$-periodic fractional Josephson effect in the absence of local mixing for a junction of proximitized quantum spin Hall edges~\cite{Zhang14}).
Notice that after a $2\pi$ evolution, the local fermion parity in the junction changes.
We track $8$ states, $4$ with even parity and $4$ with odd parity, and these states all intersect at $\phi = \pi$.

However, adding interactions
\begin{align}
H_\text{int}& = w_1 (i\gamma_{L1}\gamma_{R1})(i\gamma_{L2}\gamma_{R2}) \notag
\\ &~~+ w_2 (i\gamma_{L1}\gamma_{R1})(i\gamma_{L3}\gamma_{R3})
\end{align}
will shift the different bands up or down.
Instead of crossing at $\pi$, many crossings are now shifted away, and so symmetry-breaking perturbations may be added that open up avoided crossings.
Not all crossings are avoided;  recall that even parity states get mapped to odd parity states and vice versa.
Crossings between these states are protected by fermion parity; we recover the $4\pi$ periodic Josephson effect.

\item $m=4$.
As predicted by Ref.~\onlinecite{Fidkowski09}, adding interactions to a system with $8$ Majoranas makes the system trivial; the term
\begin{align}
H_\text{int}& = w_1 (i\gamma_{L1}\gamma_{R1})(i\gamma_{L2}\gamma_{R2}) \notag
\\ &~~+ w_2 (i\gamma_{L1}\gamma_{R1})(i\gamma_{L3}\gamma_{R3}) \notag
\\ &~~+ w_3 (i\gamma_{L1}\gamma_{R1})(i\gamma_{L4}\gamma_{R4}) \notag
\\ &~~+ w_4 (i\gamma_{L1}\gamma_{L2})(i\gamma_{L3}\gamma_{L4}) \notag
\end{align} completely removes any degeneracy at the crossing while respecting time reversal.
The Josephson effect is $2\pi$ periodic.
\end{itemize}

%%%
\bibliography{FJE}

\end{document}